\documentclass[bibyear]{aa}

\usepackage{graphicx}
\usepackage{txfonts}

\usepackage{natbib,twoopt}
\usepackage[breaklinks=true]{hyperref}
\hypersetup{
    colorlinks=true,
    linkcolor=red,
    citecolor=blue,
    filecolor=magenta,      
    urlcolor=cyan,
}  
\bibpunct{(}{)}{;}{a}{}{,} 

\begin{document}

\title{The application of the Quark-Hadron Chiral Parity-Doublet Model \\ to neutron star matter}

\author{A. Mukherjee
          \inst{1}
          \and
          S. Schramm
          \inst{2}
          \and
          J. Steinheimer
          \inst{3}
          \and
          V. Dexheimer
          \inst{4}}
          
\institute{Frankfurt Institute for Advanced Studies,
D-60438, Frankfurt am Main, Germany\\
         \email{mukherjee@fias.uni-frankfurt.de}
         \and
         \email{schramm@fias.uni-frankfurt.de}
         \and
         \email{steinheimer@fias.uni-frankfurt.de}
         \and
         Department of Physics, Kent State University, Kent, OH 44242, USA\\
         \email{vdexheim@kent.edu}}         


\abstract{}
{The Quark-Hadron Chiral Parity-Doublet model (Q$\chi$P) is applied to calculate compact star properties in the presence of a deconfinement phase transition.} 
{Within this model, a consistent description of nuclear matter properties, chiral symmetry restoration, and a transition from hadronic to quark and gluonic degrees of freedom is possible within one unified approach.} 
{We find that the equation of state obtained is consistent with recent perturbative quantum chromodynamics (QCD) results and is able to accommodate observational constraints of massive and small neutron stars. Furthermore, we show that important features of the equation of state, such as the symmetry energy and its slope, are well within their observational constraints.}
{ }

\titlerunning{Q$\chi$P modelling of neutron-star matter}
\authorrunning{Mukherjee et al.}
\maketitle

\section{Introduction}
\label{intro}

Understanding strong interaction physics at extreme conditions of temperature and/or density is  a central topic of current theoretical and experimental nuclear research. Recently, lattice quantum chromodynamics (QCD) calculations have established that chiral symmetry restoration proceeds as a smooth crossover at vanishing baryo-chemical potential ($\mu_{\rm B} = 0$). The pseudo-critical temperature, defined by the maximal change of the order parameter of the chiral transition (the chiral condensate) was found to be approximately $T_{\rm c}= 155 \pm 10$ MeV (\cite{Borsanyi:2010cj,Bazavov:2010sb}). Since lattice QCD results cannot be directly extended to finite chemical potential (\cite{Fodor:2001pe,Fodor:2004nz,deForcrand:2008vr,Endrodi:2011gv}), one has to rely on expansions of $\mu_{\rm B}=0$ lattice data in powers of chemical potential, imaginary chemical potential extensions (\cite{deForcrand:2008vr}), reweighing techniques (\cite{Fodor:2001pe,Fodor:2004nz}), complex Langevin approaches (\cite{Sexty:2013ica}), functional renormalization group approaches (\cite{Pawlowski:2005xe,Berges:2000ew,Schaefer:2006sr}), or effective models (see, e.g., Refs. \cite{Mishustin:1993ub,Heide:1993yz} for early ideas and \cite{Papazoglou:1997uw,Papazoglou:1998vr,Tsubakihara:2009zb} for hadronic models), to estimate the phase structure of QCD at large baryon number densities.

The study of ultra-relativistic heavy-ion collisions  provides  a crucial tool to  obtain information about  high temperature and, depending on the beam energy, low to high baryon number densities. In contrast, the properties of compact stars, such as neutron or hybrid stars, might  lead to a better understanding of extremely dense, but relatively cold matter.  Finally, the environment of proto-neutron stars and binary neutron star mergers could help to bridge these two regimes of density and temperature.
In order to theoretically investigate the properties of strongly interacting matter in all of these  environments, one has to employ a model approach containing hadrons and quarks in a comprehensive manner. After having analyzed (isospin-symmetric) heavy-ion-collision matter within the aforementioned approach (\cite{Mukherjee:2016nhb}) based on an extension of the SU(3) parity-doublet model of \cite{Steinheimer:2011ea},
we use the same model to investigate the behavior of hybrid stars  in the current paper. As an important outcome, we see that not only does this approach lead to a good description of the ground state properties of nuclear matter, but such an approach also leads to heavy hybrid stars with relatively smaller radii. In addition, the quarks do not feature a strong repulsive vector interaction, which would otherwise lead to conflicts with lattice QCD data (\cite{steinheimer2014lattice,Steinheimer:2010sp,kunihiro1991quark}).

The paper is organized as follows: after introducing the basic model equations and discussing isospin-symmetric matter in Sec. \ref{model}, we  determine the behavior of isospin-asymmetric matter in Sec. \ref{isoam}. Sec. \ref{neutr} contains results for  the properties of compact stars, which is followed by our conclusions and an outlook.
\section{The model}
\label{model}

In the Quark-Hadron Chiral Parity Doublet model (Q$\chi$P), an explicit mass term for baryons, which preserves chiral symmetry, is introduced in the Lagrangian. In this case, the signature for chiral symmetry restoration is the degeneracy of the usual baryons and their respective negative-parity partner states. This is different from the standard chiral picture, where this degeneracy occurs only for essentially massless nucleons. Recent results from lattice QCD, in fact, indicate that the mass of the ground state baryons undergo only a very small change and, as chiral symmetry is restored, mainly the parity partners undergo a significant mass shift (\cite{Aarts:2017rrl}).

Positive and negative parity states of the baryons
can be grouped in doublets $N = (N^+,N^-)$ as discussed in \cite{PhysRevD.39.2805,Hatsuda:1988mv}. The three-flavor extension of this approach was first presented in  \cite{Nemoto:1998um}, the application to compact stars can be found in \cite{Dexheimer:2008cv,Dexheimer:2007tn,Dexheimer:2012eu} and the application to the Q$\chi$P model was first introduced in \cite{Steinheimer:2011ea}.
To describe the interaction potential, one constructs SU(3)-invariant terms in the Lagrangian including the meson-baryon and meson-meson self-interaction terms outlined in \cite{Papazoglou:1997uw}. Taking into account the scalar and vector condensates in mean-field approximation, the resulting baryon-Lagrangian (${\cal L_B}$) reads (\cite{Steinheimer:2011ea})

\begin{eqnarray}
{\cal L_B} &=& \sum_i (\bar{B_\text{i}} i {\partial\!\!\!/} B_\text{i})+ \sum_i  \left(\bar{B_\text{i}} m^*_\text{i} B_\text{i} \right) \nonumber \\ &+&
\sum_i  \left(\bar{B_\text{i}} \gamma_\mu (g_{\omega \text{i}} \omega^\mu +
g_{\rho \text{i}} \rho^\mu + g_{\phi \text{i}} \phi^\mu) B_\text{i} \right) ~,
\label{lagrangian2}
\end{eqnarray}
summing over the states of the baryon octet, where all the $g_{\phi {\rm i}}$s are set to zero for this study. As already mentioned, this model allows for a significant bare mass term $m_0$ of the baryons. The effective masses of the baryons then follow as
\begin{eqnarray}
m^*_{\text{i}\pm} = \sqrt{ \left[ (g^{(1)}_{\sigma \text{i}} \sigma + g^{(1)}_{\zeta \text{i}}  \zeta )^2 + (m_0+n_\text{s} m_\text{s})^2 \right]}\nonumber \\
\pm g^{(2)}_{\sigma \text{i}} \sigma \pm g^{(2)}_{\zeta \text{i}} \zeta ~,
\label{effmass}
\end{eqnarray}
where $g^\text{(j)}_\text{i}$ are the coupling constants of the baryons with the non-strange ($\sigma$) and strange ($\zeta$) scalar fields. In addition, there is an SU(3) symmetry-breaking mass term proportional to the strangeness, $n_\text{s}$, of the respective baryon.
Parity-doublet models allow for two different scalar coupling terms with $i~(\text{or } j)=1 \text{ and } 2$, which lead to a different dependence of the mass of the parity-partners when the value of the chiral condensate changes. For simplicity, we assume equal mass difference of the various baryons and their parity partners in vacuum, by setting $g^{(2)}_{\zeta {\rm i}} = 0$ and $g^{(2)}_{\sigma {\rm i}} = (m_{\rm n+}-m_{\rm n-})/2\sigma_0$, where $\sigma_0$ is the vacuum expectation value of the chiral field $\sigma$.

The scalar meson interaction between the baryons drives the spontaneous breaking of the chiral symmetry. While preserving the relevant symmetries, the interaction can be written in terms of the SU(3) invariants
$I_2 = (\sigma^2+\zeta^2) ,~ I_4 = -(\sigma^4/2+\zeta^4) $ and $I_6 = (\sigma^6 + 4\zeta^6) $ as:\begin{equation}
V = V_0 + \frac{1}{2} k_0 I_2 - k_1 I_2^2 - k_2 I_4 + k_6 I_6 ~,
\label{veff}
\end{equation}
where $V_0$ is fixed by demanding that the potential vanishes in vacuum. As has been pointed out in \cite{Horowitz:2002mb,dexheimer2015reconciling}, and references therein, and discussed in general in  \cite{Schramm:2002xa}, a coupling term between 
$\omega$ and $\rho$ meson leads to a reduced value of the slope parameter of the symmetry energy. Without this coupling, the slope parameter is close to a value of $100$ MeV, which is rather large compared to current estimates. We introduce such a term in the model as
\begin{equation}
V_{\omega\rho} =  \beta \omega^2 \rho^2~~~.
\label{vor}
\end{equation}
For the sake of simplicity, we did not add this term in an SU(3) invariant way, although it is possible to do so in principle. This is reasonable because the strange-vector field necessary for the invariance, $\phi$, is effectively zero, as no hyperons occur in the system at relevant densities.
The parameters seen above are summarized in Table \ref{modpar}.

\begin{table}[b]
\begin{center}
\begin{tabular}{ | c | c | c | }
\hline
 \rule{0pt}{3ex}$k_0$ & $k_1$ & $k_2$ \\ 
 $(242.61 \text{ MeV})^2$ & 4.818 & -23.357 \\
 \hline
 \rule{0pt}{3ex}$k_6$ & $\varepsilon$ & $g_\sigma^{1,1}$ \\
 $(0.276)^6 \text{ MeV}^{-2}$ & $(75.98 \text{ MeV})^4$ & \hspace{1.5mm} -8.239296 \hspace{1.5mm} \\
 \hline
  \rule{0pt}{3ex}$g_{\rm N \rho}$ & $\delta m_{\rm q}$ & $\delta m_{\rm s}$ \\ 
 4.55 & 6 & 150\\
 \hline
 \rule{0pt}{3ex} $g_\sigma^{1,8}$ & $\alpha_\sigma^{1}$ & $g_{\text{N}\omega}$ \\ 
 -0.936200 & 2.435059 & 5.45\\
 \hline
 \rule{0pt}{3ex}$g_{{\rm q}\sigma}$ & $g_{{\rm s}\zeta}$ & $\beta$ \\ 
 2.5 & 2.5 & 900\\
 \hline
\end{tabular}
\vspace{+1.0mm}
\caption{Model parameters. The SU(3) couplings $g_\sigma^{1,1}$, $g_\sigma^{1,8}$ and  $\alpha_\sigma^{1}$ determine the baryonic coupling strengths $g_{\sigma {\rm i}}^{(1)}$ and  $g_{\zeta {\rm i}}^{(1)}$ as in \cite{Papazoglou:1998vr}.}
\label{modpar}
\end{center}
\end{table}

As the quarks also couple to the scalar fields, their masses are partly generated by the scalar mesons except for an explicit
mass term ($\delta m_\text{q}$ for up and down quarks, and $\delta m_\text{s}$ for the strange quarks) and $m_{0\text{q}}$ as follows:
\begin{eqnarray}
~m_\text{q}^*&=&g_{\text{q}\sigma}\sigma+\delta m_\text{q} + m_{0\text{q}}~,\nonumber\\
~m_\text{s}^*&=&g_{\text{s}\zeta}\zeta+\delta m_\text{s} + m_{0\text{q}}~.
\label{mqms}
\end{eqnarray}
The mass parameter for the quarks is $m_{0\text{q}}= 165$ MeV. This additional mass term can be understood as a coupling of the quarks to the dilaton field (gluon condensate). Since it is known (\cite{Sasaki:2011sd}) that the dilaton field slowly vanishes (much slower than the chiral condensate), the quark mass can still be significantly larger than the current mass $\delta m_q$ around the transition line. Eventually, if coupled to the dilaton field, the $m_{0\text{q}}$ should also asymptotically vanish in the de-confined phase. However, for simplicity, we leave it constant in this work.
Given such a mass term, the quarks do not appear in the nuclear ground state, which would be a clearly nonphysical result. This also allows us to set the vector-type repulsive interaction strength of the quarks to zero. A non-zero vector interaction strength would lead to a massive deviation of the quark number susceptibilities from lattice data, as has been observed in different mean field studies (\cite{kunihiro1991quark,Ferroni:2010xf,Steinheimer:2010sp,steinheimer2014lattice}).

\begin{figure}[t]
\includegraphics[width=0.5\textwidth]{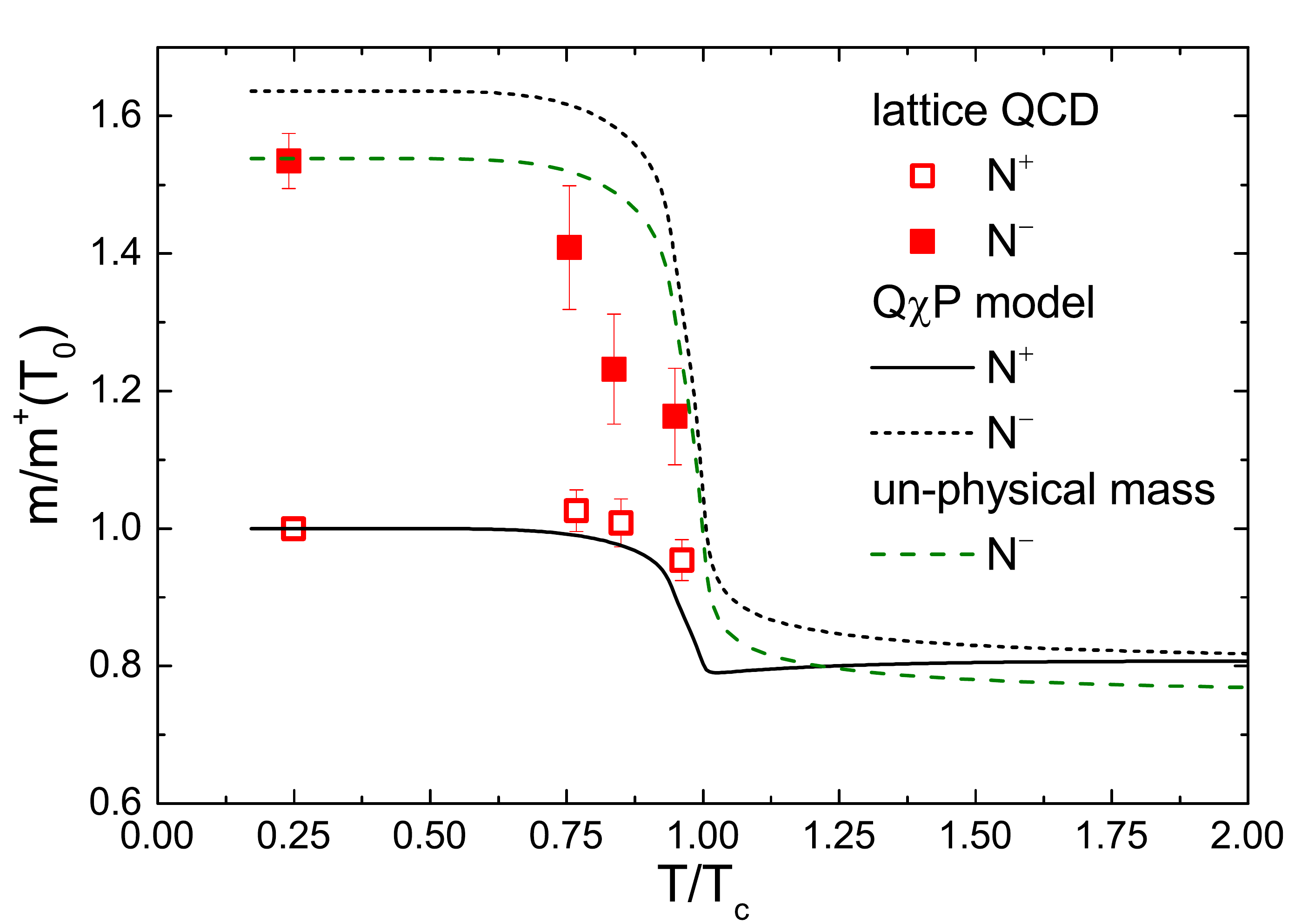}
\caption{ Mass of the nucleon ground state $\rm N^+$ and its parity partner $\rm N^-$, normalized to the $\rm N^+$ mass, as function of normalized temperature for $\mu_{\rm B}=0$ and isospin symmetric matter. We compare results from the Q$\chi$P model with lattice QCD data (\cite{Aarts:2017rrl}). For comparison, we also include Q$\chi$P results (green, dashed line), where $\rm N^-$ is normalized by the lattice QCD nucleon mass, which in the cited study is still unphysically large.}
\label{massT}
\end{figure}

To include a transition from a hadronic to a deconfined quark phase in the model, the  thermal contribution of the quarks is added to the grand canonical potential of the model, analogously to the way it is performed in PNJL type models (\cite{Fukushima:2003fw,Ratti:2005jh}) as follows:

\begin{equation}
\Omega_\text{quark}=-T \sum_{{\rm i}\in Q}{\frac{\gamma_\text{i}}{(2 \pi)^3}\int{d^3k \ln\left(1+\Phi \exp{\frac{E_\text{i}^*-\mu_\text{i}}{T}}\right)}}
\end{equation}
and
\begin{equation}
\noindent
\Omega_{\overline{\text{quark}}}=-T \sum_{\text{i}\in Q}{\frac{\gamma_\text{i}}{(2 \pi)^3}\int{d^3k \ln\left(1+\Phi^* \exp{\frac{E_\text{i}^*+\mu_\text{i}}{T}}\right)}}~.
\end{equation}
The sums run over all quark flavors, where $\gamma_\text{i}$ is the corresponding degeneracy factor, $E_\text{i}^*=\sqrt{m_\text{i}^{*2}+p^2}$ the energy, and $\mu_\text{i}$ the chemical potential of the quark.

At low temperatures and densities, the quarks are confined by the Polyakov loop potential (\cite{Ratti:2005jh}),
\begin{eqnarray}
        U&=&-\frac12 a(T)\Phi\Phi^*\nonumber\\
         &+&b(T)\ln[1-6\Phi\Phi^*+4(\Phi^3\Phi^{*3})-3(\Phi\Phi^*)^2],
\end{eqnarray}
 where $a(T)=a_0 T^4+a_1 T_0 T^3+a_2 T_0^2 T^2$, $b(T)=b_3 T_0^3 T$.\\
The parameters $a_0, a_1, a_2$, and $b_3$ are initially fixed, as in \cite{Ratti:2005jh}, by demanding a first order phase transition in the pure gauge sector at $T_0=270$ MeV. The Stefan-Boltzmann limit of a gas of gluons is reached in the limit $T\rightarrow\infty$.

Finally, in order to slowly remove the hadrons from the system as deconfinement is realized, we introduce an excluded volume description for the hadrons. The parameter $v_\text{i}$ is the volume excluded by a particle of species $i$, where we only distinguish between baryons, mesons, and quarks. Consequently, $v_\text{i}$ assume three values written as
\begin{align*}
&v_\text{quark} =~0~, \nonumber \\
&v_\text{baryon} =~v~, \nonumber \\
&v_\text{meson} =~v/a~, \nonumber
\end{align*}
where $a$ is a number larger than one. In our calculations, we choose the value $a$ = 8. This treatment eventually modifies the effective chemical potential of all the hadrons, causing these hadrons to be suppressed once the quarks and gluons contribute to the thermodynamic potential. As a result, we obtain a naturally smooth transition from a hadronic to a quark-dominated system. A detailed description of the excluded volume prescription can be found in  \cite{Mukherjee:2016nhb,Steinheimer:2010ib,Steinheimer:2010sp}.

\begin{figure}[t]
\includegraphics[width=0.35\textwidth, angle = 270]{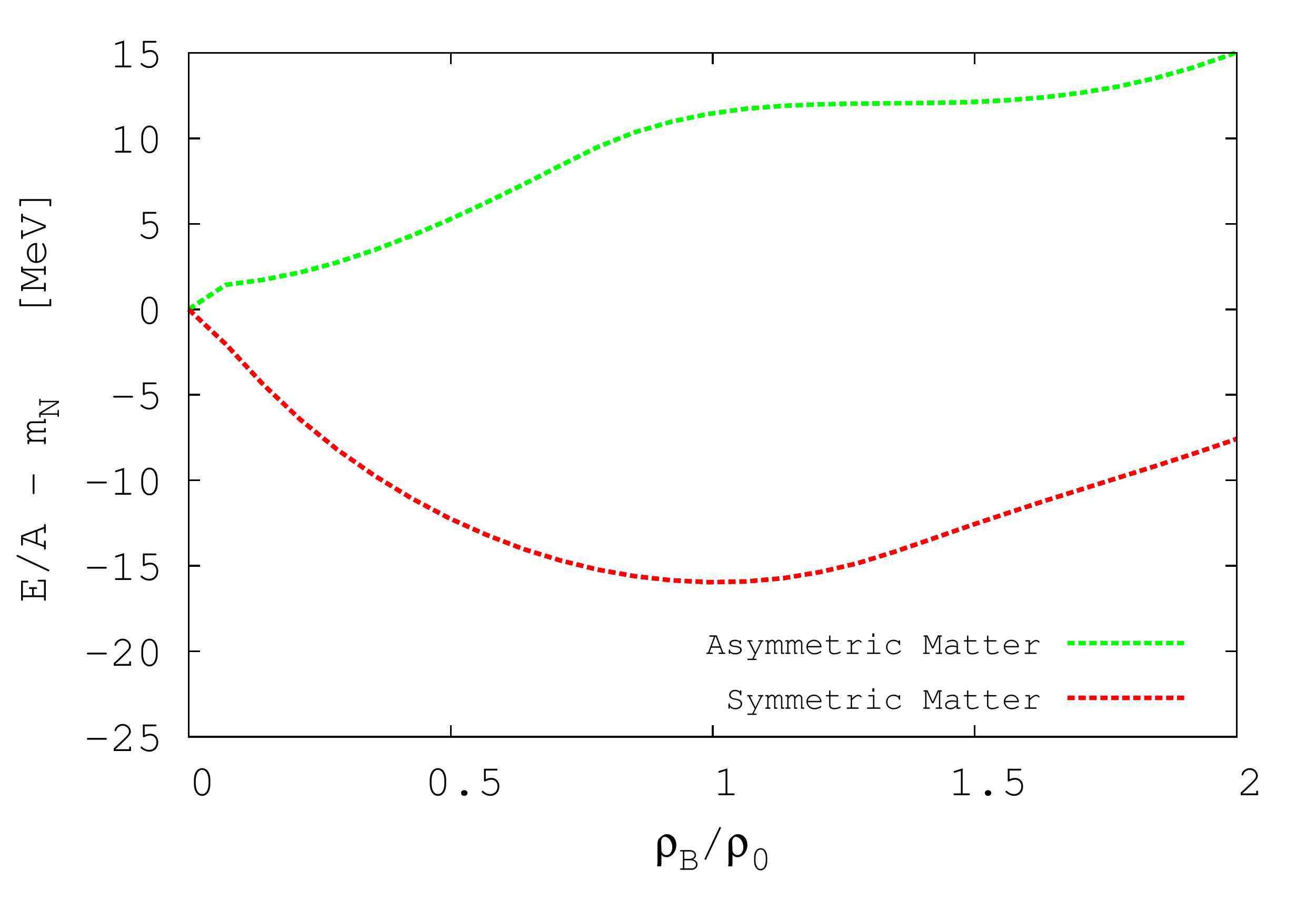}
\caption{ Binding energies for symmetric and asymmetric nuclear matter, as functions of the baryon number density, which is normalized to nuclear matter saturation density $\rho_0$.}
\label{bind}
\end{figure}

Results for this model, regarding properties of isospin symmetric matter, have already been shown in \cite{Mukherjee:2016nhb}. 
A particular feature of the parity doublet model is that the nucleon and its negative parity partner suffer a different change in their mass from the melting chiral condensate. To put our model in context of recent lattice QCD calculations we show, in 
Fig. \ref{massT}, a comparison of the mass of the nucleon and its parity partner with lattice QCD data (\cite{Aarts:2017rrl}) as a function of temperature, at $\mu_B=0$. Even though the vacuum mass of the nucleons in the lattice QCD calculations is still off its physical value, the temperature dependence shows a remarkable similarity with our results. This result indeed supports a basic assumption of the Q$\chi$P model, where chiral symmetry restoration is observed in a mass degeneracy for hadrons and their parity partner and not in a complete absence of mass. 

When we studied the model properties at $T=0$ we found that, for the parameters used in the previous paper and in this work, nuclear saturation properties are well described. Even the nuclear (in)compressibility ($\kappa$) was found to be 267 MeV, which is well in line with experimental observations. The red dashed line in Fig. \ref{bind} shows the resulting binding energy per baryon for infinite and isospin-symmetric nuclear matter, as a function of the normalized net-baryon-number density $\rho_{\rm B}/\rho_0$, where $\rho_0$ ($=0.142$ fm$^{-3}$) is the nuclear saturation density of the model.

\section{Isospin$-$asymmetric matter}
\label{isoam}

As we investigate neutron stars, i.e., highly isospin-asymmetric matter, we first determined the basic isospin-dependent coefficients around nuclear matter saturation density. To this end, we calculated the value of the  isospin-asymmetry energy $S_{\rm v}$, given by

\begin{equation}
S_{\rm v} = \frac{1}{8} \left[ \frac{d^2(\varepsilon/\rho_{\rm B})}{d (I_3/B)^2} \right]_{\rho_{\rm B}=\rho_0}~,
\label{symmdef}
\end{equation}
where $\varepsilon$ is the energy density, $\rho_{\rm B}$ the baryon number density, $I_3$ the isospin $3$-component, and $B$ the net baryon number. The density dependence of $S_{\rm v}$ is usually parametrized by the slope parameter
\begin{equation}
L = 3\rho_0 \left[ \frac{dS}{d\rho_{\rm B}} \right]_{\rho_{\rm B}=\rho_0}~.
\label{slopedef}
\end{equation}
\noindent
Using our model, we obtain $S_{\rm v} = 30.02$ MeV and $L = 56.86$ MeV, which are in agreement with ranges of $L$ and $S_{\rm v}$, from various experiments and analyses in, for example, \cite{Lattimer:2012xj}.

\begin{figure}[t]
\includegraphics[width=0.35\textwidth, angle = 270]{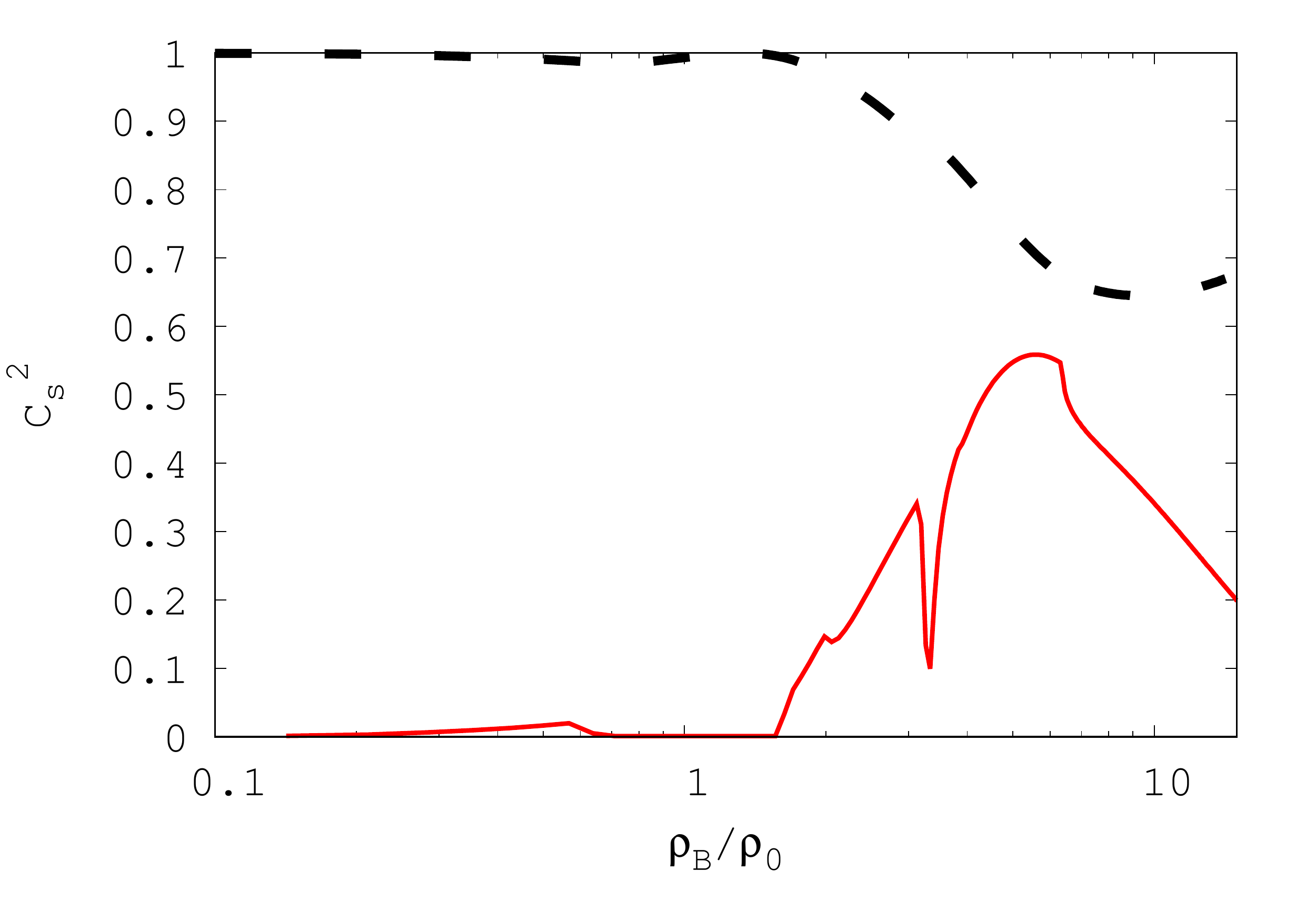}
\caption{ Speed of sound squared as a function of normalized baryon number density (on a logarithmic scale), with the dashed, black line representing the kinetic-theory-bound on $c_{\rm s}^2$ (cf. \cite{PhysRevC.95.045801}).}
\label{cs2}
\end{figure}

\begin{figure}[t]
\includegraphics[width=0.5\textwidth, angle = 0]{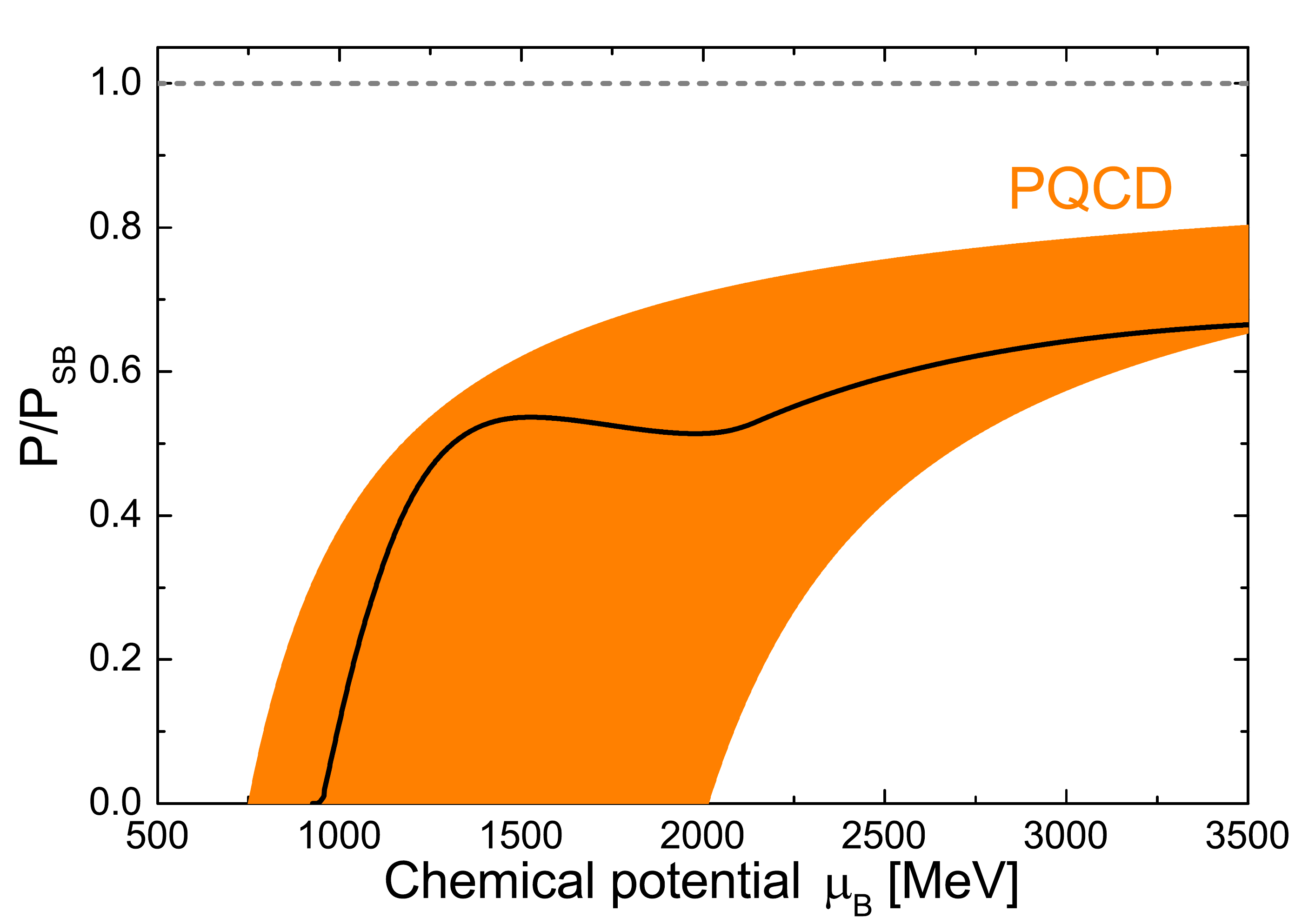}
\caption{ Comparison between the pressure, normalized to the Stefan-Boltzmann pressure, $P/P_{\rm SB}$ obtained from our model and PQCD calculations from \cite{Fraga:2013qra}.}
\label{prat}
\end{figure}

The binding energy per baryon of asymmetric star matter is also shown in Fig. \ref{bind}. 
In this case, the energy is determined self-consistently by the imposition of electric charge neutrality and chemical equilibrium, including free, charged leptons. 
In addition, we show the square of the speed of sound for star matter in Fig. \ref{cs2}. It is calculated from our equation of state as 
\begin{equation}
c_{\rm s}^2 = \left.{dP/d\varepsilon}\right|_{T=0} \ ,
\end{equation}
where $P$ is the pressure and $\varepsilon$ the energy density. 
The sharp decrease in the speed of sound around $\rho_{\rm B} \approx 3 \rho_0$ signals the appearance of the parity partner, $N^{*0}$, of the neutron, as it starts to be populated. The smaller sharp decrease in the speed of sound just before $\rho_{\rm B} \approx 2 \rho_{0}$ signals the appearance of the down quarks, although only a few quarks contribute to the particle cocktail at low density. Fig. \ref{cs2} also shows that, in our model, the speed of sound never crosses the boundary established by kinetic theory (\cite{PhysRevC.95.045801}),

\begin{equation}
\frac{c_{\rm s}^2}{c} = \frac{\varepsilon - P/3}{\varepsilon+P}
\label{ktbound}
.\end{equation}
\noindent
In addition, for very large densities, our speed of sound remains around $\sqrt{1/3}$, as expected. 

The pressure of star matter, divided by the Stefan-Boltzmann pressure (ideal-gas limit), as a function of baryo-chemical potential is shown in Fig. \ref{prat}. We compare our results with star-matter, perturbative-QCD (PQCD) calculations at zero temperature from \cite{Fraga:2013qra}, which can be considered a constraint on the high density QCD equation of state (EoS; \cite{Kurkela:2014vha}). Our EoS falls inside the band in Fig. \ref{prat}, which represents their uncertainty estimates. One should note, however, that the agreement of our model with the PQCD result gets worse for very large values of chemical potential. This is because we have assumed the quark mass parameter ($m_{0 \rm {q}}$ in Eqn. (\ref{mqms})) to remain constant for all densities. In reality, we expect that, as the dilaton field melts slowly at large values of chemical potential, the quark mass also slowly approaches the current quark mass value, i.e., the quark mass parameter should vanish. Thus, for high values of the chemical potential, our model shows a rise in pressure that is too slow near the Stefan-Boltzmann limit.

\begin{figure}[t!]
\includegraphics[width=0.35\textwidth, angle = 270]{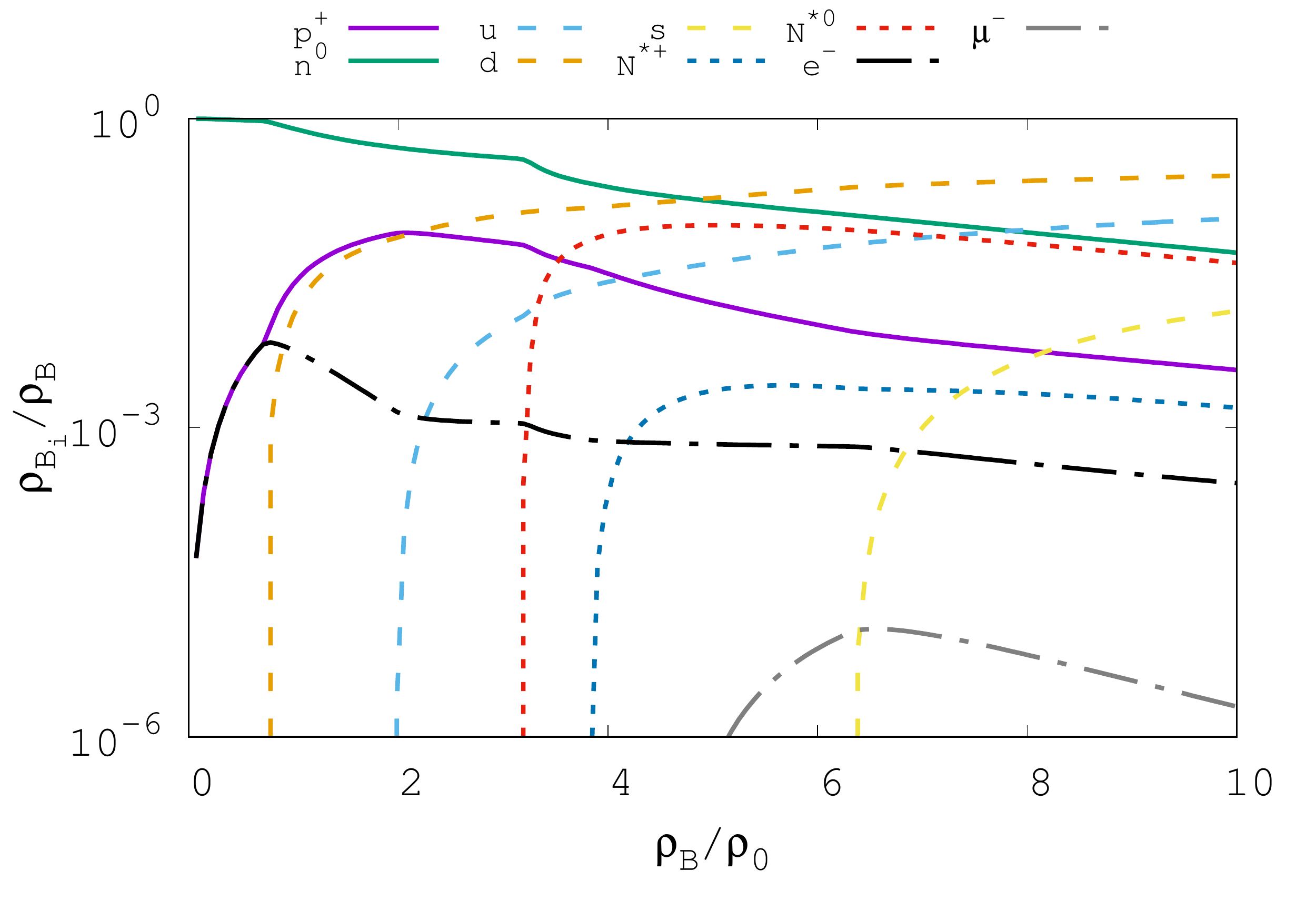}
\caption{ Normalized baryon-number densities of various particle species as functions of the normalized baryon-number density.}
\label{dense}
\end{figure}

In order to better understand the chemical composition of our asymmetric EoS, we determined the corresponding particle populations. Fig. \ref{dense} shows the number densities of various particle species normalized to the total baryon number, where quark number densities are divided by 3. At low densities, as expected, we only observe neutrons. An increase in density is followed by the appearance of protons, leptons and, soon after, down-quarks. Later, the up-quarks appear, followed by the chiral partner of the neutrons. The latter appears rather suddenly and causes the rapid decrease of the speed of sound mentioned earlier (cf. Fig. \ref{cs2}). Finally, the chiral partner for the protons, and afterwards the strange quarks, appear. Although the hyperons are included in the model, they are completely absent from the particle cocktail shown in Fig. \ref{dense}. The chiral partners of the nucleons have lower masses than the hyperons and owing to the crossover formalism, quarks can also appear very early. In addition, the hyperons are suppressed by the appearance of the other light quark states through the excluded volume formalism. Eventually the strange quarks appear in the cocktail, however only at much higher densities.

\begin{figure}[t!]
\includegraphics[width=0.5\textwidth, angle = 0]{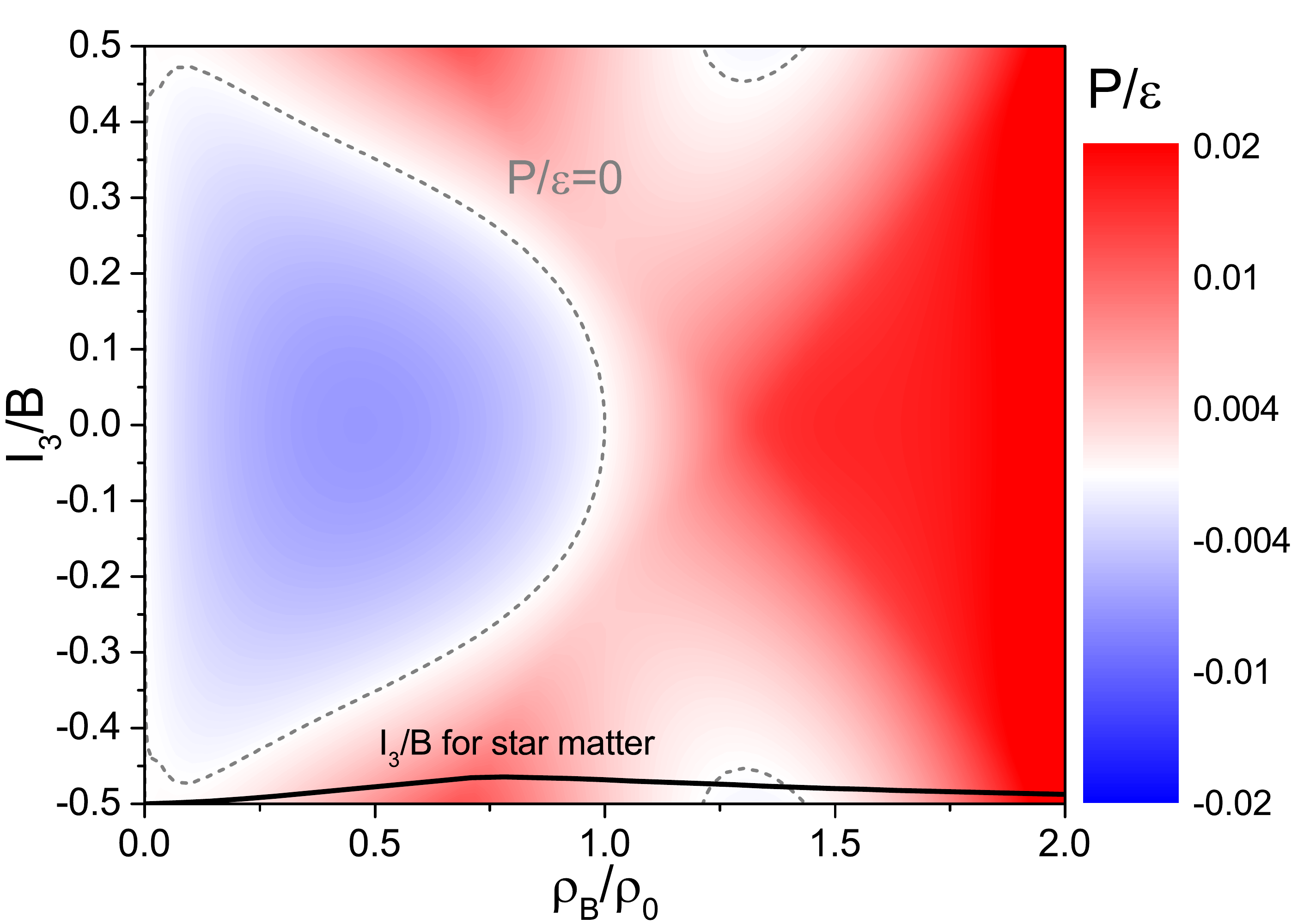}
\caption{ Contour plot of the ratio $P/\varepsilon$ on a normalized $I_3-\rho_{\rm B}$ plane, showing regions of stability in red and those of instability in blue. The bold, black line represents the isospin-per-baryon for charge-neutral and chemically equilibrated matter .}
\label{pe}
\end{figure}

As we have already mentioned, the isospin asymmetry of charge neutral and chemically equilibrated matter is self-consistently determined. In this case, we show in Fig.~\ref{pe} how the pressure to energy density ratio $P/\varepsilon$ changes as a function of isospin per baryon, which is defined as
\begin{equation}
\frac{I_3}{B}=\sum_{\rm i} \frac{(I_{3})_{\rm i} \rho_{\rm i}}{\rho_{\rm B}} ~,
\end{equation}
and the normalized baryon density. The colors in the figure show regions where the pressure is positive (red) and negative (blue). All the unstable and metastable states of the nuclear liquid-gas transition fall into the blue region at small baryon number densities. We also observe small regions, both at large and small values of isospin-per-baryon, where the pressure decreases as function of density, or in other words, where the speed of sound becomes imaginary and matter becomes mechanically unstable. This region corresponds to the spinodal region of a first-order phase transition, which appears only for large isospin asymmetries. The bold, black line corresponds to the EoS of neutron star matter, where the isospin per baryon is fixed by condition of beta-equilibrium.

\section{Neutron stars}
\label{neutr}

\begin{figure}[t]
\includegraphics[width=0.5\textwidth, angle = 0]{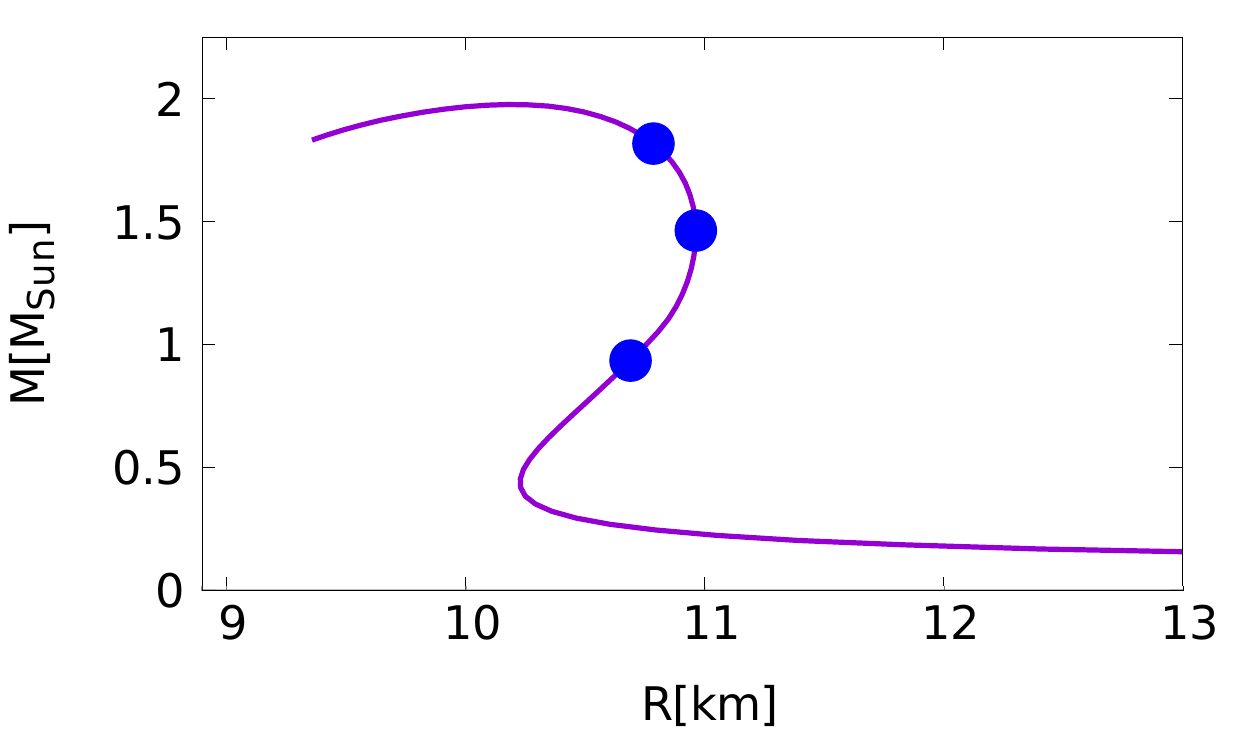}
\caption{ Mass-radius diagram. The blue dots indicate stars with a fraction of $20\%$, $25\%,$ and $30\%$ of the baryon mass coming from quarks.}
\label{mr}
\end{figure}

In order to describe neutron stars we make use of a Maxwell construction around the first-order phase transition to avoid thermodynamical instabilities, i.e., we guarantee that the pressure increases as a function of energy density in our EoS. We also add a standard result for the crust to our EoS, originally calculated by Baym, Pethick, and Sutherland (\cite{Baym:1971pw}). The mass-radius diagram for the resulting compact stars, as shown in Fig.~\ref{mr}, is determined using the Tolman-Oppenheimer-Volkoff (TOV) equations (\cite{PhysRev.55.374,PhysRev.55.364}) for a range of central pressures. The most massive star of the family has a mass of $1.98$ ${\rm M}_{\odot}$ (${\rm M}_{\odot} = $ the solar mass) and a radius of $10.25$ km. The canonical $1.4$ ${\rm M}_{\odot}$ star has a radius of $11.10$ km. 
This radius value, which is small for models of hybrid or other exotic matter, is in agreement with a number of observational studies, particularly of low-mass X-ray binaries that point to small neutron star radii in the range of about 9 km to 11 km (\cite{Guillot:2013wu,Guillot:2010zp,Ozel:2016oaf,Ozel:2015gia}). Since most of the stellar cores we reproduce contain some amount of quarks, we choose to mark the stars from the family that contain $20\%$, $25\%,$ and $30\%$ of baryon mass coming from quarks (blue dots in Fig.~\ref{mr}). For the most massive star of the family $35\%$ of its baryon mass is generated by quark matter.

\begin{figure}[t]
\vspace*{-0.8cm}
\includegraphics[width=0.5\textwidth, angle = 0]{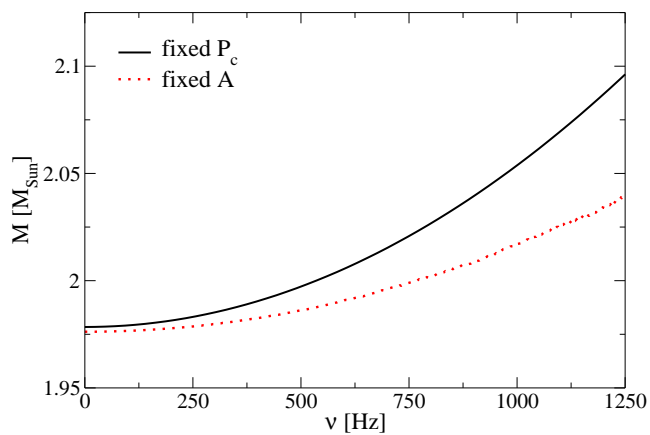}
\caption{ Mass of the most massive star of the family as a function of rotational frequency. In two cases, we keep a fixed central pressure or number of baryons in the star.} 
\label{rot}
\end{figure}

When we include rotation effects, Fig.~\ref{rot} shows how the stellar maximum mass increases as a function of rotational frequency in two cases, keeping a fixed central pressure or the number of baryons in the star. In the second case, we describe the evolution of an isolated star, as the frequency of rotation decreases over time, which has a Kepler frequency of 1606 Hz. Here, we considered monopole and quadrupole corrections to the metric due to the rotation, as was derived in \cite{Glendenning:1993di}. The higher the rotational frequency, the more massive and larger the stars become. The increase in mass of the most massive star of the family is about $5\%$ for a star with fixed baryon number rotating at its Kepler frequency (compared to a non-rotating star). In a previous publication we have shown that this kind of calculation differs by about $1\%$ from full general relativity results from \cite{Stergioulas:1994ea}.

Usually, the effect of rotation in hybrid stars is to suppress their quark phase (cf. \cite{Wei:2017qha} for a recent discussion on the topic). A phase transition to deconfined matter can only take place when
heavy stars spin down and their central densities increase. In our case, the situation is different because the quarks occupy a fraction of almost all stars, heavy and light. We have quarks present in almost all stars that rotate with any allowed frequency and only their fraction (compared to hadrons) increases as stars spin down.

\begin{figure}[t]
\vspace*{-0.7cm}
\includegraphics[width=0.5\textwidth, angle = 0,trim={3.4cm 0 0 0},clip]{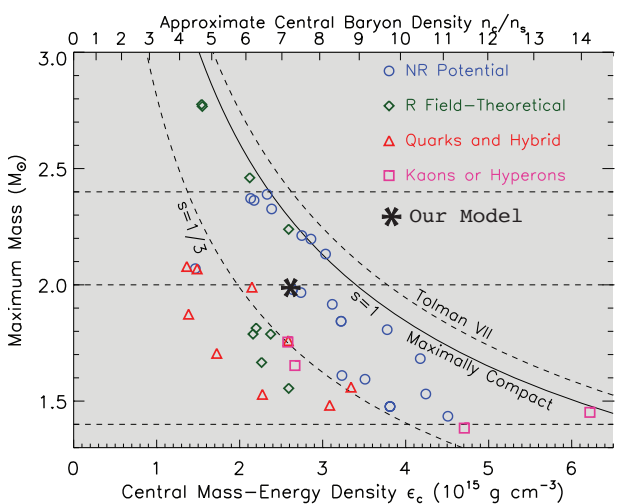}
\vspace*{-0.8cm}
\caption{ Compactness (stellar mass vs. central density) of the most massive star generated by our equation of state. Other equations of state shown are calculated using nonrelativistic models, relativistic models, models containing quarks, and models containing strange hadrons. The figure has been adapted from \cite{Lattimer:2004sa,Lattimer:2012nd}.}
\label{comp}
\end{figure}

The term ``compactness'' refers to how packed together matter is in a star. Our Q$\chi$P equation of state is shown in Fig.~\ref{comp} together with other equations of state calculated using nonrelativistic models, relativistic models, models containing quarks, and models containing strange hadrons. This figure was adapted from \cite{Lattimer:2004sa, Lattimer:2012nd}. It is interesting to see that the star we generate is very compact, and is more compact than all the massive exotic stars shown in the figure. Naturally, the maximum-mass star we reproduce is between the line that represent EoS with constant speeds-of-sound, equal to $\sqrt{1/3}$ or 1.

Concerning star cooling, our EoS does not allow the hadronic direct Urca process. This is the case because, although a large fraction of the star core contains nucleons and their parity partners, there are not enough electrons to complete the reactions. For a more detailed study of the role of chiral partners in neutron star cooling, see \cite{Lattimer:2012nd}. For quarks, we  assumed that all flavors are paired and as such the quark direct Urca process is heavily suppressed (\cite{Blaschke:2005dc,Page:2005fq,Alford:2004zr,Negreiros:2010tf}). The absence of the direct Urca process is a large advantage of our EoS, as it prevents the enhanced cooling of heavy stars, as discussed in \cite{Page:2004fy,negreiros2013impact}.

\section{Conclusions and outlook}
\label{conc}

We presented results  for  isospin asymmetric matter and compact star  properties within the Quark-Hadron Chiral Parity-Doublet model (Q$\chi$P). The model produces 2 ${\rm M}_{\odot}$ hybrid stars (without quark-vector interactions), with a  large quark fraction of about 30 percent, as quark degrees of freedom begin to be populated at low densities. Also, because of that, the hyperons do not appear in the star. However, the parity-doublet partners of the nucleons are present.

The radius of the reproduced canonical 1.4 ${\rm M}_{\odot}$ star is very small, i.e., about 11 km; this is  in accordance with the radius estimates obtained from studies of low-mass X-ray binaries (cf. \cite{Guillot:2010zp}). This behavior is also reflected in the large compactness of the reproduced maximum mass star. Additionally, extending the EoS to high chemical potential, it meets the band of pressure values  obtained in PQCD studies. At extreme values of $\mu_{\rm B}$ ( $> 3500$ MeV), the model underestimates the pressure mainly due to the remaining large bare mass term of the quarks.

An important result of our work is that we have at hand a single model for the description of hybrid stars with a hadronic and a quark phase. The properties of the EoS are different from most simpler models, which usually incorporate the phase transition from a hadronic to a quark phase through an artificial construction. Thus, we presented a hybrid EoS that leads to more compact stars and still allows for a large quark fraction, while not forbidding the appearance of hyperons. 

The application of the Q$\chi$P-model EoS to dynamic simulations for heavy ion collisions can be used to study observables for the QCD phase transition in iso-spin symmetric matter. At the same time, numerical studies of neutron-star mergers can be conducted with the same model EoS in a consistent manner. This enables us to study nuclear matter in very different environments and in systems of vastly different scales using a single EoS.

\section*{Acknowledgements}
The authors would like to thank G. Aarts and M. Hanauske for helpful discussions and M. Strickland for carefully reading the manuscript. The authors also acknowledge the support from NewCompStar COST Action MP1304, from GSI, HIC for FAIR, and from BMBF. The computational resources were provided by the Frankfurt Centre for Scientific Computing (CSC) and Center for Nuclear Research (Kent State University). In addition, the authors acknowledge the insights and comments of the referee who took the time to review this paper.
 
\bibliographystyle{aa} 
\bibliography{bibnew.bib} 

\end{document}